\documentclass[superscriptaddress, preprintnumbers,amsmath,amssymb]{revtex4}

\usepackage{graphicx}
\usepackage{bm}
\include{epsf}

\begin{document}


\title{Cooperative peer-to-peer multiagent based systems}


\author{L. F. Caram}
\affiliation{Laboratorio de Redes y Sistemas M\'{o}viles, FI-UBA. \\Av. Paseo Col\'on 850, Buenos Aires, C1063ACV, Argentina\\
Email: fcaram@fi.uba.ar\\ 
}
\author{C. F. Caiafa}
\affiliation{Instituto Argentino de Radioastronom\'{i}a (CCT La Plata, CONICET), \\C.C.5, 
Villa Elisa, Buenos Aires, 1894, Argentina\\
Email: ccaiafa@iar.unlp.edu.ar\\ 
}
\author{M. Ausloos}
\affiliation{School of Management, University of Leicester, 
 University Road, Leicester, LE1 7RH,  UK \\Email: ma683@le.ac.uk\\
Group for Research on Applications of Physics in Economy and Sociology (GRAPES), 
\\  rue de la Belle Jardiniere, 483/021,  B-4031 Liege Angleur, Belgium  \\Email: marcel.ausloos@ulg.ac.be\\  e-Humanities group, Royal Netherlands Academy of Arts and Sciences, \\Joan Muyskenweg 25, 1096 CJ Amsterdam, The Netherlands\\Email: marcel.ausloos@ehumanities.knaw.nl\\ 
}
\author{A. N. Proto$^\dagger$}
\affiliation{Laboratorio de Sistemas Complejos, FI-UBA. \\Av. Paseo Col\'on 850, 
Buenos Aires, C1063ACV, Argentina,\\ $ ^\dagger$deceased\\
} 


\date{\today}

\begin{abstract}
A  multiagent based model for a  system of cooperative agents aiming at growth is proposed. This is based on a set of generalized Verhulst-Lotka-Volterra differential equations. In this study, strong cooperation is allowed  among agents having similar sizes, and weak cooperation if agent have markedly different  ``sizes",  thus establishing a peer-to-peer modulated interaction scheme.
A rigorous analysis of the stable configurations is presented first examining the fixed points of the system, next determining their stability as a function of the model parameters.
It is found that  the agents are self-organizing into clusters. 
Furthermore, it is demonstrated that, depending on parameter values, multiple stable configurations can coexist. It occurs that only one of them always  emerges   with probability close to one, because its associated attractor dominates over the rest. This is shown through numerical integrations and simulations,after analytic developments.
In contrast to  the competitive case,  agents  are able  to increase their capacity beyond the no-interaction case  limit. In other words, when some collaborative partnership among a relatively small number of partners takes place,   all agents act in good faith prioritizing the common good, whence receiving a mutual benefit allowing them to surpass their   capacity.
\end{abstract}


\maketitle

\section{Introduction}

It is usually accepted that the fittest survive through natural selection \cite{Darwin}. The former biological  feature has been extended to  social and moral concepts. It should be recalled  that Darwin's conclusion pertained to  the  preservation of favored races in the struggle for life, when species  performed in a given  environment. However, the question of species competition $or$ cooperation  among themselves for survival is another matter.

Generally speaking, species social systems are in fact classified in {\it competitive}, {\it cooperative} or {\it mixed} type, depending on the set of interactions existing among agents. For example, in Nature, ants exhibit a typical cooperative behavior \cite{dorigo2004ant,ANTS}, on the other side, in economy, companies sharing a customer market usually behave as natural competitors \cite{fehr1999theory}. However,  collaboration with  competitors might be a winning strategy \cite{hamel1989collaborate,rey2013cooperation,luo2006cross,hauert2006cooperation,doi:10.1142/S0219525912500592,doi:10.1142/S0219525913500367}.

Thus,  fundamental, beside   moral or economic,  questions  can be raised about sets of interacting agents exhibiting emergence of a self-organizing collective behavior, not resulting from the existence of a``central controller" \cite{boccara2004modeling,bar1997dynamics,Knyazeva,porter2000location}, but due to their own interactions with the other agents.

Several simulation and analytic studies can be found  in the literature on such systems, called {\it prey-predator models}.  However, for such  systems, the agent interacting processes  can be  also  modeled by using a set of ordinary differential equations (ODEs), for example, following a Lotka-Volterra (LV) model \cite{boccara2004modeling,Lotka,Volterra}.  The time evolution or {\it dynamics} of the system can be displayed along a continuous time axis, rather than at discrete time points, in simulation work. 

For example, in \cite{Huberman,Huberman2} the LV model was used to model the competition among websites using constant and equal interactions among all agents. It was found that two distinctive behaviors are possible: {\it winner takes all} and {\it sharing the market}. In \cite{Yanhui}, this model was modified by introducing a non-constant and linear interaction which allows the emergence of the {\it rich gets richer} behavior. Moreover, in \cite{Caram}, a competitive $non-linear$ interaction was considered  leading to a stratification or clustering  of agents, as often  observed in economic life. 

In this work, a cooperative scenario between agents, rather than a competitive one, is presented, based on a similar set of LV model differential equations. Particularly,   how the ``size" of the each agent increases (or not) is studied depending on the cooperation with the other agents. When modelling  such a socio-economic multi agent system, the ``size" is understood as something similar to  the market share, \cite{Huberman,Yanhui,Caram}. In this line of thought,  the $n$ interacting agents are all needing some common resources within a general environment. Here below, an interaction function is introduced, which allows agents to cooperate in a selective way, i.e. the interaction is strong between those agents with similar or equal sizes. On the other hand, a weaker cooperative  interaction  is between agents if they have very different sizes. As a side way argument, it can be considered that such a same-size-cooperative rule occurs in sport competition. For example, the main (soccer) teams share the best players in order to remain at the top.

Here, a simple symmetric model is considered, i.e. by assuming that the strength of cooperation is reciprocal between two cooperating agents. It is shown that this model allows for unvealing the main features of this interesting type of systems. More sophisticated interactions including, for example, asymmetry in the cooperation could be addressed in the future as an extension of this wok (see section \ref{sec:concl}). 
Therefore, here our cooperative system has two parameters,  $\sigma$ and $K$, (see equation (\ref{eq:model}) in Section \ref{sec:model}):\begin{itemize} \item  (a) $\sigma$ scales the difference between agent sizes, \item (b) $K$ defines the kind of scenario (cooperative or not) and also controls the amplitude of  the interactions. \end{itemize}

The main purpose of this paper is to analyze the cooperative scenario and demonstrate that a synergy is established leading to a situation in which every interacting agent is benefited from the group. This is the point that makes the difference with other previous works \cite{Huberman,Yanhui,Caram}.

This paper is organized as follows: in Section~\ref{sec:model}, the mathematical model is described; in Section~\ref{sec:analysis3}, the cooperative scenario is studied in detail; in Section~\ref{sec:analysisFP}, the fixed points are searched and the  $K$ parameter range effect is analysed; in Section~\ref{sec:simul}, simulations and results for a case of ten agents are shown; finally, in Section~\ref{sec:concl}, the main conclusions are outlined.  

\section{\label{sec:model}The model}
Let  $n$ agents be sharing some common resource; when an  agent is able to get some portion of the common resource, its size increases,  but if  losing a portion of its resource  its size decreases. In the model, essentially based on the well known prey-predator model, the interaction parameter is not assumed to be a constant: it  is supposed to depend on the difference between agent sizes. This fact implies  that sizes do dynamically change in time, expectedly producing a highly complex dynamics, - due to a  feedback phenomenon.
Mathematically, the model   is  based on   Verhulst evolution equations  \cite{Verhulst1,Verhulst2}
\begin{equation}\label{Veq}
\dot{s}_{i}=\alpha_{i}s_{i}\left(\beta_{i}-s_{i}\right) ,
\end{equation}  
   and on the generalized Lotka-Volterra   evolution equations  \cite{Lotka,Volterra}
\begin{equation}\label{LVeq}
\dot{s}_{i}=\alpha_{i}s_{i}\left(\beta_{i}-s_{i}\right)-{\sum\limits_{i\neq j}}\gamma_{ij}s_{i}s_{j} , \mbox{ \ \ \ \ \ \ for \ \ \ \ }i\neq j ,   \mbox{ \ \ \ \ \ \ and \ \ \ \ }i=1,\dots,n
\end{equation}  
where $s_{i}$ is the size of agent $i$, $\alpha_{i}$ is the agent's growth rate, $\beta_{i}$  is the agent's maximum  capacity and $\gamma_{ij}$ is a constant coefficient determining the interaction between $s_{i}$ and $s_{j}$. Eq.(\ref{LVeq}) is easily generalized to read
\begin{equation}\label{protoeq}
\dot{s}_{i}=\alpha_{i}s_{i}\left(\beta_{i}-s_{i}\right)-{\sum\limits_{i\neq j}}\gamma\left(s_{i},s_{j}\right)s_{i}s_{j}, \mbox{ \ \ \ \ \ \ for \ \ \ \ }i=1,\dots,n,
\end{equation}%
where   $\gamma\left(s_{i},s_{j}\right)$ is the interaction between  agent $s_{i}$ and agent $s_{j}$, here this matrix is composed by dynamic parameters that result from the difference between both agent sizes in relation. There is no time lag delay. Therefore, each agent size $s_{i}(t)$ represents a portion of common resource that the agent is able to get at a given time $t$. When the interaction $\gamma\left(s_{i},s_{j}\right)\neq 0$,   a feedback situation is present; if $\gamma=0$ the system is reduced to the basic uncoupled LV / Kolmogorov prey-predator model \cite{Lotka,Volterra}. In this case, each agent size  grows according to its  particular rate $\alpha_{i}$, up to   its  maximum capacity $\beta_{i}$ or  its possible maximum size, as it happens in the population dynamics model of Verhulst \cite{Verhulst1,Verhulst2}. 

Thereafter, the interaction function $\gamma\left(s_{i},s_{j}\right)$ is  supposed to be non linear, symmetric, and monotonically decreasing with distance from its centre \cite{Caram,broomhead1988radial}
\begin{equation}
\gamma\left(s_{i},s_{j}\right)=K\exp\left[-\left(\frac{s_{i}-s_{j}}{\sigma}\right)^{2}\right],   \label{eq:gamma}
\end{equation}%
where $\sigma >0$, is a positive (kernel bandwidth) parameter that controls or scales the size similarity i.e. it regulates the difference in size of agents and determines the interaction levels. On the other hand, $K$ determines the type of scenario: $K>0$ (competitive case,  as studied in  \cite{Caram}), or $K<0$ (cooperative case), as studied here below. The absolute value $|K|$ defines the amplitude of the interaction. Therefore, the dynamics is dominated by the interaction which can be ``strong" or ``weak" depending on the sizes of agents. It seems obvious that when a big agent is interacting with a small one, the intensity of  their interaction is  weak or almost null, since their ``size distance" is large. {\it A contrario}, when two agents with the same or similar sizes are interacting, the intensity of their interaction  can be very strong. It is easy to see from Eq.(\ref{eq:gamma})  that $\gamma\left(s_{i},s_{j}\right)$ varies between $0$ and $K$. 

It is  not too difficult to conserve a full set of  different $\alpha_{i} $ and $\beta_{i}$, characterising each agent. However, the writing is much more heavy if doing so.  In order to remain  within the purpose of this paper, it is advantageous to  consider that all agents have the same basic dynamics properties. This is equivalent to rescaling the various sizes and the time scales. Thus, thereafter, let $\alpha_{i}=1$ and $\beta_{i}=1$. The whole model  becomes:
\begin{equation}
\dot{s}_{i}=s_{i}\left(1-s_{i}\right)-{\sum\limits_{i\neq j}}K\exp \left[{-\left(\frac{s_{i}-s_{j}}{\sigma }\right)^{2}}\right] s_{i}s_{j}.   
 \label{eq:model}
\end{equation}

\section{Detailed analysis of the cooperative scenario}\label{sec:analysis3}
A cooperative scenario occurs when $K$ is negative in  Eq.(\ref{eq:model}), i.e. creating a positive feedback. Instead of stabilizing the system,  this leads to  very unstable  and complex behaviors. Therefore, the various interesting values of $K$, must be chosen carefully. It will be seen in section \ref{subsec:Trivialfp}, that  this choice depends on the total number of agents  which are cooperating. This analysis will show that there is a quite limited range of $K$ values such that the model system reaches  a stable behavior in the steady state.

\begin{figure}
\begin{center}
\includegraphics[width=14cm] {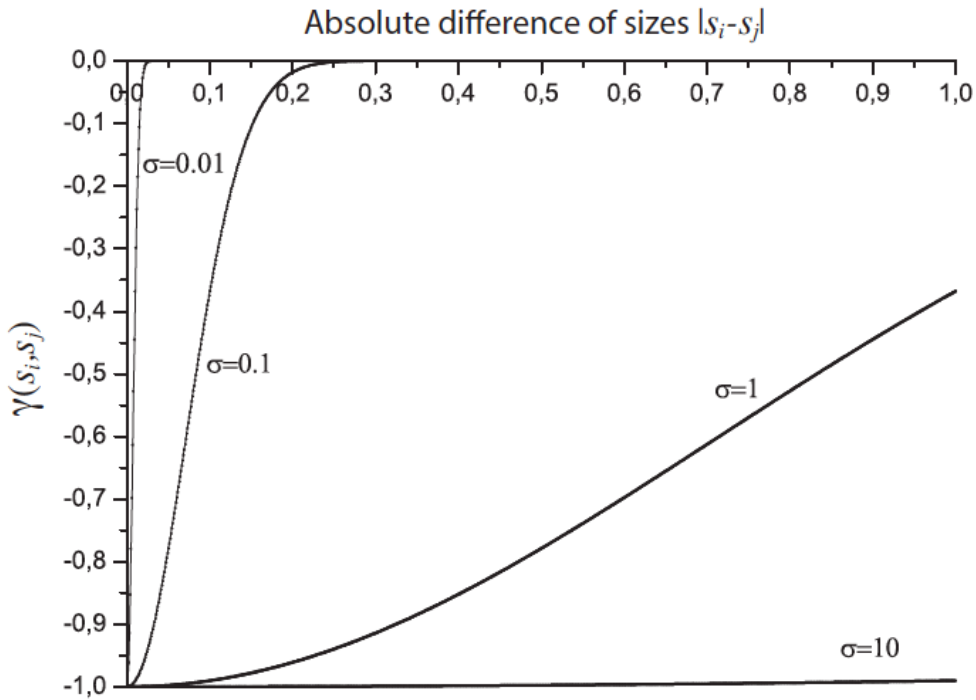}
\caption{Interaction function $\gamma$ vs. absolute difference of sizes $|s_{i}-s_{j}|$, for different scaling similarity parameter $\sigma$ values and for $K=-1$.}
\end{center}
\label{fig:F1}
\end{figure}

First, it is obvious that cooperation is  at its maximum and equal to $K$,  when $s_{i}=s_{j}$, because of  the definition of  $\gamma\left(s_{i},s_{j}\right)$. The possible variation range of $\gamma\left(s_{i},s_{j}\right)$ is shown in
Figure 1,  as a function of the difference in sizes, for different values of $\sigma$. As can be seen in this figure, when the (absolute) difference in  sizes increases, the cooperation decreases: $\gamma \left(s_{i},s_{j}\right)$  becomes  less negative;   its possible values all within the range $-K<\gamma\left(s_{i},s_{j}\right)<0$. It is clear that, for example when $\sigma=10$ and $\left|s_{i}-s_{j}\right|<1$, the cooperation is almost constant and equal to $\gamma_{ij}\left(s_{i},s_{j}\right)=-K$. When $\sigma\rightarrow\ 0$, the interaction function looks like an inverted Dirac distribution, at the origin.

In the following Section \ref{sec:analysisFP}, it is shown how a narrow and limited range of $K$ values, for different number of agents, leads the system to  stable configurations in the steady stationary state. The emergence of different behaviors will be also shown, i.e.  the formation of clusters of agents whose sizes are larger than their limit,  i.e. its maximum capacity ($\beta_{i}=1$)  in absence of interaction. More interestingly, the coexistence of several stable configurations is proved analytically and validated by numerical simulations.

\section{Analysis of fixed points and range of parameter $K$}\label{sec:analysisFP}

By definition, a fixed point (FP) is a point in the phase space where all the time derivatives are zero, i.e.,%
\begin{equation}
\dot{s}_{i}=0, \mbox{ \ \ \ \ \ \ for \ \ \ \ } i=1,\dots,n.  \label{eq:condicionPF}
\end{equation} 

For the  stability  analysis associated to every fixed point, one looks at the eigenvalues of the Jacobian matrix $J$ evaluated at the corresponding FP. It is rather easily derived   that  the Jacobian matrix  elements  are:%
\begin{equation}
\left[ J\right] _{\left( i,k\right) }=\frac{\partial \dot{s}_{i}}{\partial s_{k}}=\left\{ 
\begin{array}{ll}
1-2s_{i}-{\sum\limits_{i\neq j}}s_{j}\gamma \left(s_{i},s_{j}\right) \left[1-\frac{2}{\sigma ^{2}}s_{i}\left(s_{i}-s_{j}\right) \right],  & \mbox{\ \ for \  }k=i, \\ 
-s_{i}\gamma \left(s_{i},s_{k}\right) \left[1+\frac{2}{\sigma^{2}}s_{k}\left(s_{i}-s_{k}\right) \right],  & \mbox{\ \ for \  }k\neq i.%
\end{array}%
\right.   \label{eq:Jacobian}
\end{equation}

\subsection{Trivial fixed points for  an arbitrary number $n$ of agents}\label{subsec:Trivialfp}

In this section, the existence of fixed points and their stability analysis are presented  when the   number of agents allows some analytical work. 

From equations (\ref{eq:model}) and (\ref{eq:condicionPF}), at least three trivial  FP  can be  detected:
\begin{itemize}
\item
(I) $s_{i}=0$, \mbox{ \ \ \ \ \ \ \ \ \ \ \ \ \ \ \ \ \ \ \ \ for \ \ \ \ }$i=1,\dots,n$ ,\\ i.e.
 all agents have zero size;
\item
(II) $s_{i}=1$ and $s_{j}=0$, \mbox{ \ \ \ \ for every \ \ \ \ }$j\neq i$ ,\\ i.e. all agents have zero size, except one;
\item
(III) $s_{i}=b$, \mbox{  \ \ \ \ \ \ \ \ \ \ \ \ \ \ \ \ \ \ \ \ for \ \ \ \ }$i=1,\dots,n$ , \\ i.e.
 all agents  have  the same size $b$.
\end{itemize}

In addition to the   type (I), (II) and (III) FP, there are many other points that verify the condition of a fixed point. These points are found by numerically seeking the roots of the non-linear Eq.  (\ref{eq:condicionPF}); see Sect. \ref{nontrivialFP}.

If the Jacobian matrix is   evaluated  at the   type (I) FP, from Eq.(\ref{eq:Jacobian})     the identity matrix is obtained;   all eigenvalues are equal to one $\left(\lambda_{i}=1\right)$. Therefore, it is an unstable fixed point.

Next, evaluating Eq.(\ref{eq:Jacobian}) at the second type (II) of fixed points,  e.g.  for the case  $s_{1}=1$ and $s_{2}=s_{3}=,\dots,=s_{n}=0$,  one gets %
\begin{equation*}
J=\left[ 
\begin{array}{ccccc}
-1 & -a & -a & \dots & -a \\ 
0 & 1-a & 0 & \dots & 0 \\ 
0 & 0 & 1-a & \dots & 0 \\ 
: & : & : & \dots & : \\ 
0 & 0 & 0 & \dots & 1-a%
\end{array}%
\right], 
\end{equation*}%
with 
\begin{equation*}
a  \equiv K\exp \left(-\sigma^{-2}\right) .
\end{equation*}

It can be shown that the eigenvalues of $J$ in this case are:%
\begin{equation}
\begin{array}{l}
\lambda_{1}=-1, \\ 
\lambda_{2,3,\dots,n}=1-a=1-K\exp\left(-\sigma^{-2}\right).%
\end{array}
\label{eq:autovaloresII}
\end{equation}

From these  equations, it can be  concluded that, when $K<0$, the type (II)  FP is not stable since it has $n-1$ positive eigenvalues; this fact is neither dependent on the number of agents nor on the value of the parameter $\sigma$. 

Finally, analyzing  the stability of the  type (III) FP, one can  calculate the corresponding constant $b$  from Eq.(\ref{eq:condicionPF}) as follows:

\begin{equation*}
0=b(1-b)-(n-1)Kb^{2}=1-b-(n-1)Kb=1-b[1+(n-1)K],
\end{equation*}
Thus, 

\begin{equation*}
b=\frac{1}{1+(n-1)K}, \mbox{ \ \ \ \ \ \ for \ \ \ \ } K\neq -\frac{1}{(n-1)}.
\end{equation*}

It  can be observed  that the total quantity $n$ of cooperating  agents  which are in the system, determines the amplitude of the cooperation $K$,  as well as the final size, e.g. $b$,  which characterize the agent cluster.
Furthermore, the   Jacobian matrix, evaluated at the type III fixed point reads: 
\begin{equation*}
J=\frac{1}{1+\left(n-1\right)K}\left[ 
\begin{array}{ccccc}
-1 & -K & -K & \dots & -K \\ 
-K & -1 & -K & \dots & -K \\ 
-K & -K & -1 & \dots & -K \\ 
: & : & : & \dots & : \\ 
-K & -K & -K & \dots & -1%
\end{array}%
\right], 
\end{equation*}%
whose eigenvalues are:%
\begin{equation}
\lambda_{1,\dots,n}=\frac{K-1}{1+\left( n-1\right) K}, \mbox{ \ \  for \ \ }K\neq
-\frac{1}{\left( n-1\right)},  \label{eq:autovaloresIII}
\end{equation}

which reveals that  type III  is a stable fixed point for the  range of $K$ values.

\begin{center}
$-\frac{1}{n-1}<K<1.$
\end{center}

This equation  allows    two possible scenarios: \begin{itemize} \item (i) cooperative case: $-\frac{1}{n-1}<K<0$, and
\item  (ii) competitive case : $0<K<1$, as analysed in \cite{Caram}. 
\end{itemize}  It  should be noticed that, when the quantity $n$ of cooperating agents tends to infinity  ($n\rightarrow \infty$), the $K$ range allowing this type of stable FP  is drastically reduced. In such a case, $b \rightarrow 1$.

\subsection{Non-trivial fixed points for a small number of agents}\label{nontrivialFP}
Non trivial FP can be numerically  found  ``easily" when the number of agents is small. 
In order to illustrate the analysis,  let us examine the case of $n=5$ agents. Considering the degeneracy of several solutions,  seven possible fixed points can be  detected as representing different possible scenarios and combinations of agents grouped into clusters or levels as it is summarized as follows:  
\begin{itemize}
\item Case I (one level): five agents (5) are grouped in one cluster.

\item Cases II and III (two levels): there are two configurations called (4-1) and (3-2), i.e., either composed of a group of four agents plus one lonely agent, on one hand, or  composed by a group of three agents and a group of two agents. 

\item Cases IV and V (three levels): there are two possible configurations composed by  either a (3-1-1) or a (2-2-1), respectively.

\item Case VI (four levels): there is only one configuration, called (2-1-1-1).

\item Case VII (five levels): there is only one configuration, of course,  called (1-1-1-1-1).
\end{itemize}


The type of stability associated to each  FP  can be determined numerically by evaluating the Jacobian matrix and by computing its eigenvalues \cite{strogatz2014nonlinear}. A stable configuration, which is associated to a stable FP, corresponds to a steady state of the system, which could include clusters of agents (more than one agent in a same level). 

{\it This algebra  allows  that more than one stable configuration may coexist, depending on the parameter $\sigma$ values}. 

A technical point : the above mentioned FPs were found by searching for the roots of Eq.(\ref{eq:model}) using the Newton-Raphson (NR) algorithm for a set of randomly selected seeds \cite{Myller}. To determine the stability of the above mentioned FPs, the eigenvalues of the associated Jacobian matrices were computed using a QR type algorithm for a wide range of $\sigma$ values.  

Cases where the fixed points that emerge in steady state (stable), depending on the $\sigma$ value, are outlined in Table \ref{tab:tabla1}. However,  the stability of case (I) is found to depend only on $K$, regardless of the $\sigma$ value, while case (VII) is never stable, regardless of the values of $K$ and $\sigma$. It has  also been  found that the final configuration of the system depends on the initial conditions. After 100 iteration steps different $s_{i}$ values levels of the respective fixed points are asymptotically reached. The solution for the case (VI), i.e. (2-1-1-1), appears mostly when a random set of initial conditions is used. However, when the set of initial conditions is chosen very close to  the asymptotic values characterizing other fixed points, the  three cases (III), (3-2), (II), (4-1) and (I), (5)) emerge. This fact clearly shows that four stable configurations may coexist, i.e. for the same $\sigma $ value, there are $four$ possible solutions. Practically, this hints at the difficulty of  forecasting cluster solutions, since the  initial conditions are rarely precisely known.

\begin{figure}
\begin{center}
\includegraphics[width=16cm]{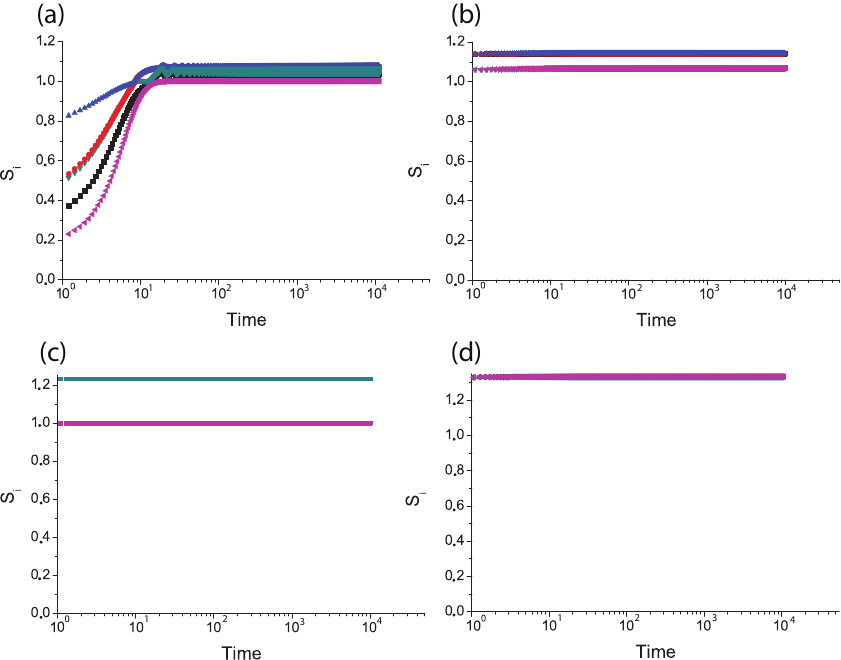}
\end{center}
\caption{Simulations of the time evolution of the size of  ($n=5$) agents for four different sets of initial conditions; all for $\sigma=0.01$. (a) CASE VI (2-1-1-1): the highest level with two agents (red and green) and the rest levels with one agent each one. Initial condition was chosen randomly in the range $[0,1]$. (b) CASE III (3-2): the highest level with three agents and the lowest level with two agents. (c) CASE II (4-1): the highest level with four agents and only one agent in the lowest level (d) CASE I (5): All agents in the same level. Initial conditions for (b), (c) and (d) were chosen very close to the corresponding FP in order to assure the convergence ($ |s_i^0 - s_i^*| < 0.01$ with $s_i^0$ is the used initial condition and $s_i^*$ is the theoretical value at the fixed point).  }
\label{fig:F2}
\end{figure}

In view of such findings, it seems of interest to display how stable configurations can be reached.  In Figure \ref{fig:F2}, the time evolution of agents, when $\sigma=0.01$, $n=5$, $K=-0.0625$, is shown;   the above remarks are emphasized by displaying  the results for four  different sets of initial conditions. Notice that the highest levels are mostly populated and the sizes always exceed $1$. This effect has always been found  in the numerical simulations, even for a different number of cooperating agents (see for example Figure \ref{fig:F5} and \ref{fig:F6}). At once, this fact shows the benefit of cooperation. When a collaborative partnership among a relatively small number of partners takes place, and all agents act in good faith prioritizing the common good, all the participants receive the same benefits.

\begin{figure}
\begin{center}
\includegraphics[width=16cm]{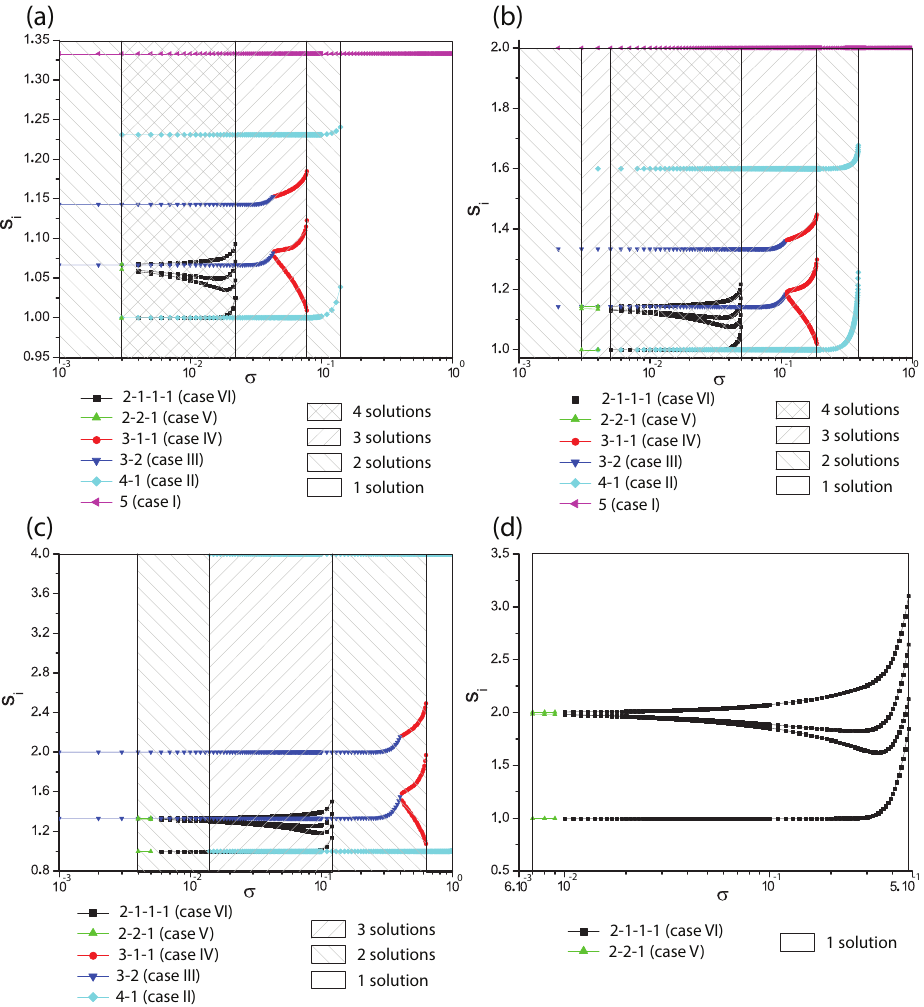}
\end{center}
\caption{Agent sizes $s_{i}$\ vs. $\sigma$ ($n=5$). Intervals in $\sigma$ with different numbers of overlapped solutions were found. (a) $K=-0.0625$ and five intervals: $ 0.001<\sigma \le 0.003$ (2 solutions), $ 0.003<\sigma \le 0.022$ (4 solutions), $ 0.022<\sigma \le 0.077$ (3 solutions), $ 0.077<\sigma \le 0.14$ (2 solutions) and $ 0.014<\sigma \le 1$ (1 solution). (b) $K=-0.125$ and six intervals: $ 0.001<\sigma \le 0.003$ (2 solutions), $ 0.003<\sigma \le 0.005$ (3 solutions), $ 0.005<\sigma \le 0.049$ (4 solutions), $ 0.049<\sigma \le 0.186$ (3 solutions), $ 0.186<\sigma \le 0.387$ (2 solutions) and $ 0.387<\sigma \le 1$ (1 solution). (c) $K= -0.25$ and five intervals: $ 0.001<\sigma \le 0.004$ (1 solution), $ 0.004<\sigma \le 0.014$ (2 solutions), $ 0.014<\sigma \le 0.12$ (3 solutions), $ 0.12<\sigma \le 0.63$ (2 solutions) and $ 0.63<\sigma \le 1$ (1 solution). (d) $K=-0.5$ and one interval with only one solution: $ 0.007<\sigma \le 0.48$.}
\label{fig:F3}
\end{figure}

\begin{figure}
\begin{center}
\includegraphics[width=16cm]{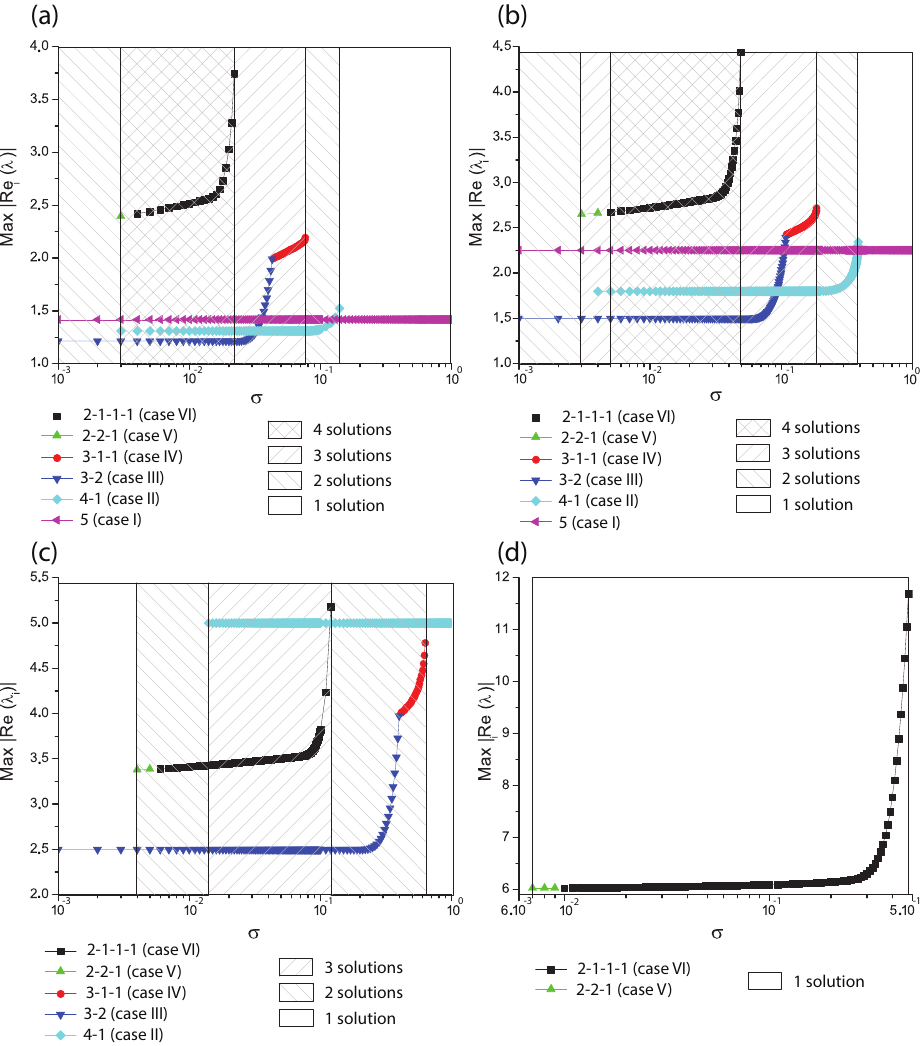}
\end{center}
\caption{Maximum absolute  eigenvalue real part of the Jacobian matrix vs. $\sigma$ value ($n=5$).  (a) $K=-0.0625$ in the interval $ 0.003<\sigma \le 0.022$ (4 solutions) the FP of type 2-1-1-1 (case VI) dominates over the rest. (b) $K=-0.125$ in the interval $ 0.005<\sigma \le 0.049$ (4 solutions) the FP of type 2-1-1-1 (case VI) dominates over the rest. (c) $K= -0.25$ in the interval $ 0.014<\sigma \le 0.12$ (3 solutions) the FP of type 4-1 (case II) dominates over the rest. (d) $K=-0.5$ in the interval $ 0.007<\sigma \le 0.48$ there is only one solution.}
\label{fig:F4}
\end{figure}

Another meaningful and enlightening type of display shows the $\sigma$ dependence of stable FPs. In Figure \ref{fig:F3}, the agents sizes that correspond to each stable FP, as a function of parameter $\sigma$, are presented for a subset of representative values of the parameter $K$. To find these FPs, the NR method was used in order to find the roots of the corresponding non linear system of equations, using 100.000 random starting points, but finally keeping only the stable FP. To evaluate numerically the eigenvalues of the Jacobian matrix at different $\sigma$ values, a QR like algorithm was used. 
From  such simulations, we confirm that there are several overlapping regions ($\sigma$ intervals) of stable configurations (see detailed explanations in the caption of Figure \ref{fig:F3}). In other words, for a given $\sigma$ value, clusters are possibly found with different asymptotic values $s_{i}$.

\subsection{Attractor strength}\label{attractor}
All eigenvalues have negative real part  in the case of a stable fixed point. Moreover, the  absolute value of the most negative eigenvalue real part is considered as determining the ``force of the attractor" or the ``speed of convergence" towards it \cite{strogatz2014nonlinear,hilborn2000chaos}. We have computed  such an absolute maximum value of the real part of eigenvalues, at each stable fixed point versus $\sigma$. This is shown in Figure \ref{fig:F4}. 

While the dynamics of the system is dominated by the FP with the highest absolute value of the real part of eigenvalue, nevertheless recall that  the precise dynamics also depends on the initial conditions. For example, Figure \ref{fig:F4} shows that the eigenvalue associated to the case (VI) (2-1-1-1) dominates  that of  the other fixed points, i.e. the probability of reaching this configuration is much higher than the other FP.  Simulations have  confirmed such a forecast.

The time evolution is non oscillatory in the case of 5 agents (see Figure \ref{fig:F2}), and it is also seen on Figures \ref{fig:F5} and \ref{fig:F6}, in the case of 10 agents, where the transition time is very short not allowing to make a visually pleasant Poincare map.

\section{Simulations and results for $n=10$}\label{sec:simul}

Even though the case of  $n    \le  5$ agents is fine enough to illustrate the main features of the model,  and their consequences (they are also summarized in the conclusion section), one might wonder whether larger systems  might  present additional features.  
Thus, we present some additional simulations results for a system with only  ten agents.  In fact, we consider that this is quite a sufficient number in order to describe  most of the economic fields, academically studied or not. 

\begin{figure}
\begin{center}
\includegraphics[width=16cm]{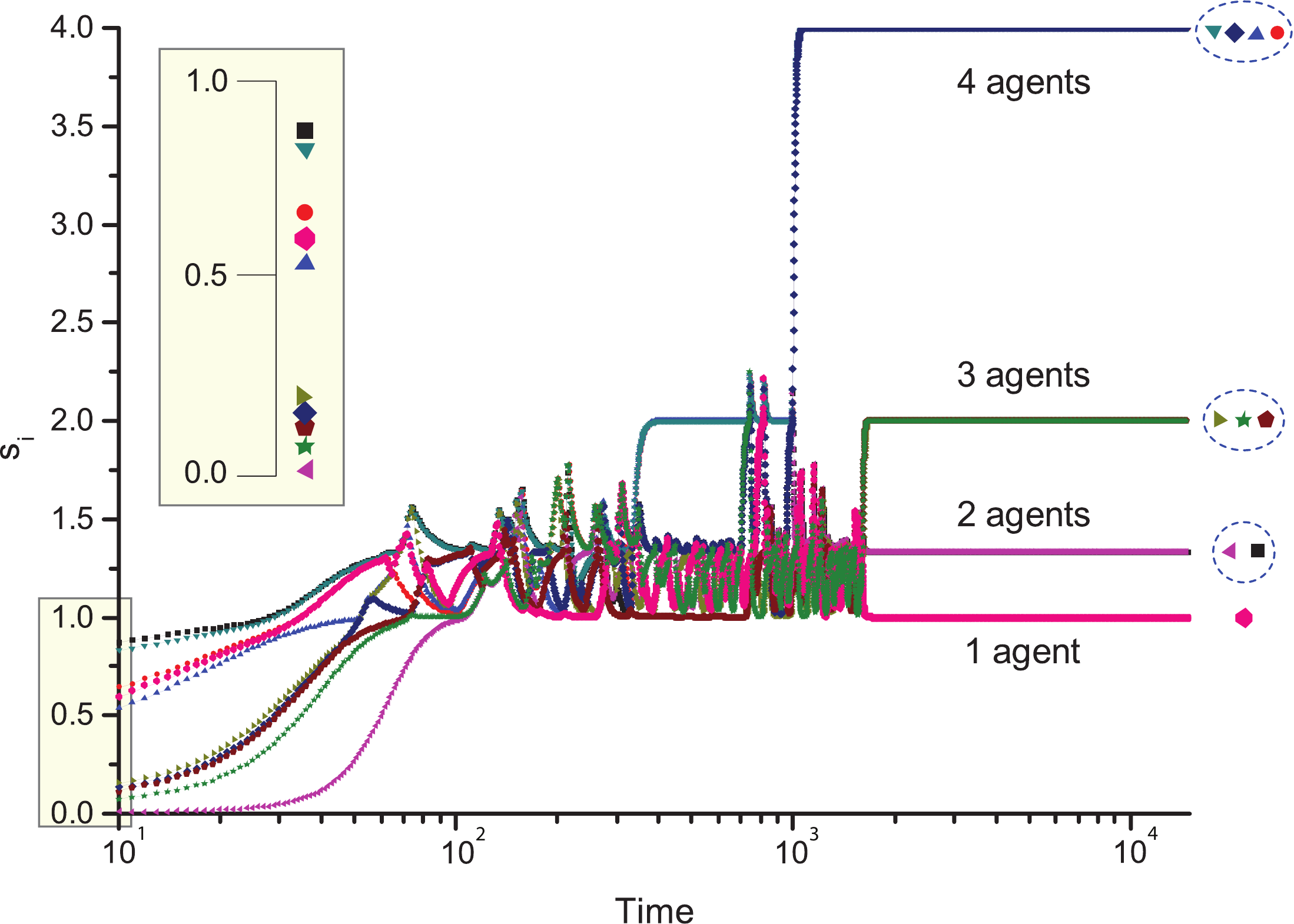}
\end{center}
\caption{Simulation example of the time evolution of the model with $n=10$, $\sigma=0.01$  and $K=-0.25$:  there are four final clusters, each containing 4, 3, 2 and 1 agents, respectively. It is highlighted that some agents with an initial small size are able to reach the highest level (see for example the agent denoted by a dark blue diamond). Conversely, some agents starting with a large size evolves to one of the lowest levels (see for example the agent denoted by a black square.)}
\label{fig:F5}
\end{figure}

\begin{figure}
\begin{center}
\includegraphics[width=16cm]{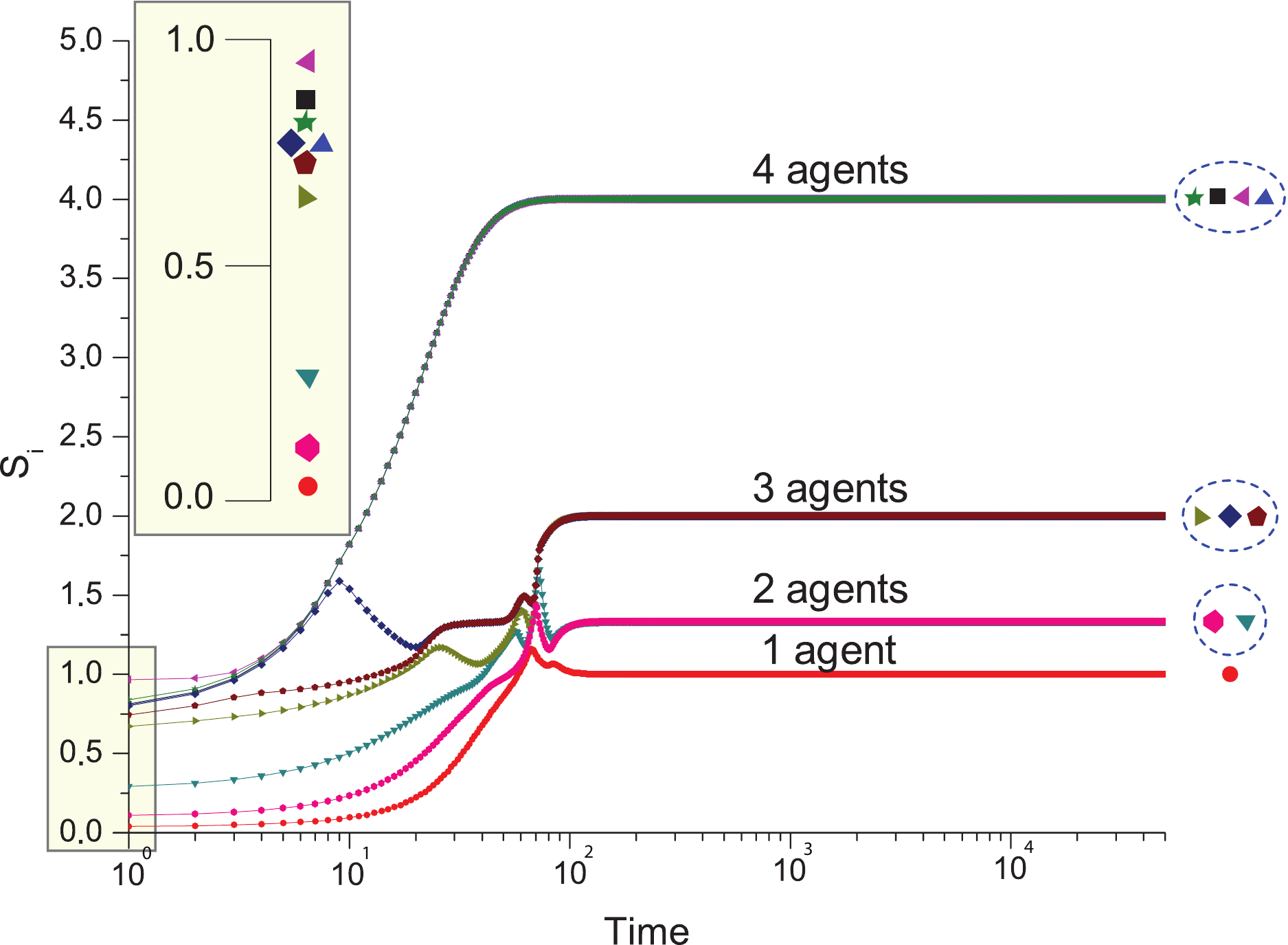}
\end{center}
\caption{Simulation example of the time evolution of the model with $n=10$, $\sigma=0.09$  and $K=-0.25$:  there are four final clusters, each containing 4, 3, 2 and 1 agents, respectively.}
\label{fig:F6}
\end{figure}

In order to span a large domain of cooperation possibilities, here $K=-0.25$ was used. The obtained behaviors are shown in Figures \ref{fig:F5} and \ref{fig:F6}, for $\sigma= 0.01$ and  $\sigma= 0.09$ respectively. These figures, illustrate that in this cooperative scenario, groups of agents are able to exceed their own size capacity limit determined by $\beta=1$ (without interaction, i.e. $\gamma\left(s_{i},s_{j}\right)=0$). It is noted that,  in the example of Fig. \ref{fig:F5}, after a chaotic transition time interval, agents reach their steady state. On the other hand, for a larger $\sigma $, see Fig. \ref{fig:F6}, i.e. for a ``long interaction size range", the agents reach their steady state faster. In these examples, the strong cooperation of four agents allows them to reach a size equal to 4. There is another group consisting of three agents whose sizes are equal to 2. Then the third group is composed of two agents of size 1.33 for each one, and finally the smaller group consists of only one agent which size is equal 1. It is also emphasized that  at this FP  the steady state was always reached, indicating that its associated probability of occurrence, after a not too long time, is very close to 1.

Therefore, this mainly simulation study on a one-shot cooperation case ($K=-0.25$) confirms the findings for a smaller number of agents under the variation of the parameters and  initial conditions, discussed in the other subsections.

\section{Conclusions and discussion}\label{sec:concl}

The study of complex systems  a very active area of research. 
The mathematical tools developed in this field are so versatile that they allow  applications to diverse kinds of problems in different research areas such as communication networks \cite{Kocarev,IEEE2011}, biology \cite{Sitharama,Bagni}, socio-economy \cite{Huberman,Yanhui}, etc.
We have considered a peer-to-peer interaction  system, but allowing for cooperation between agents rather than competition.
 
In this work, we have found different behaviors.  All lead to a structuration of the system with a remarkable group (cluster) hierarchy.  
Multiple stable solutions for a  given $\sigma$ value  can be found,  but  always one   solution  dominates.  The largest group with the largest size markedly dominates in the system. Moreover, the dominating  group occurs  more quickly when the  interaction   encompasses a wider   size range. It is interesting to note that the $\sigma$ value can be interpreted in a social or economical context as the way in which agents interact. For a small $\sigma$ value, the interaction is restricted to agents with very similar sizes and agents with different sizes do not interact with each other. On the other side, having a large $\sigma$ value allows agents with different sizes interact, which is close to the meaning of social equality. In other words, in the latter case, powerful (big) social groups are able to cooperate with weak (small) groups. In the case of an economical system, having a large $\sigma$ value may correspond to a Òfree marketÓ situation in which all agents (big and small ones) are allowed to interact each other. On the other hand, having a small $\sigma$ value may correspond to the case of having government regulations that constrain the agents to interact with only agents of similar sizes. 

Summarizing, sometimes a solution, e.g. case (VII) in Sect. \ref{nontrivialFP}  is never stable, regardless of $\sigma$, $K$ and $n$ values, while  cases are always stable for $K> \frac{-1}{n-1}$, e.g. case (I).  According to $\sigma$  ranges,  the spread of cooperation,  there are stable steady states, but sometimes   not. Thus, it is  possible to have more than one stable configuration,  ... two, three or even four,  in the  5-agent case, depending on the $\sigma$ value. Moreover, the ranges of   $\sigma$ where the  steady state solution is unique can be  also observed, (see Table \ref{tab:tabla1}).  Stressing the $K$ value order of magnitude, like    $K=-1$, case no stable solution is found, regardless of  the $\sigma$ value

 After making a large number  ($10^4$) of numerical simulations  for uniformly distributed random initial conditions, we have always obtained the solution corresponding to the eigenvalue with highest absolute value of the real part. This suggests that the probability to obtain other solutions is almost vanishing. In contrast, for non-uniformly distributed initial conditions and in particular those close to other relevant fixed points,  possible solutions  emerge. In so doing, we can claim that there is much   coherence in the model.

From a ``practical" point of view, our main  finding has been to show the power of cooperation in order to increase ``size".    When agents cooperate, they are able to triply or even quadruply  increase their size, in some sense their market share. This should be contrasted with   previous  studies in which  the competitive scenario leads to the winner takes all:    agents are clustered at sizes lower than 1, i.e. their  theoretical  capacity when behaving independently of each other \cite{Huberman,Huberman2,Caram}. Cooperation, instead, allows  for the  group of cooperative agents to configure a cluster with a characteristic level higher than the individual capacity, obtained without interactions, which is defined by the $\beta=1$ value.

The model describes approximately what is happening in society, at least in common sense expectations. Consider a few examples (i) "cooperation rather than competition between Coca-Cola and Pepsi-Cola in order to share the market and avoid many intruders; (ii) cooperation between a few co-authors in order to improve their number of publications, citations, whence h-index; (iii) in sport, cooperation (within theoretical competition) in order to win a race or a game; (iv) cooperation in the car industry in order to be the first to propose an electric car (Daimler AG, parent of Mercedes-Benz cooperate with Tesla Motors;   RenaultÐNissan Alliance has made agreements to promote emission-free mobility in France, Israel, Portugal, Denmark); (v) let us briefly mention competition AND cooperation between political parties in order to form a coalition government (when the cooperating weakest ones can overcome the top party, though a case indeed not found in our model)

It is highlighted that we have considered here only symmetric interactions allowing us to unveil the main advantages of cooperation. It is worth mentioning that in real cooperative socio-economical systems sometimes not all agents cooperate in the same way. This important characteristic of real systems could be incorporated in the model through an asymmetric interaction kernel in a future work, for example, by considering $K_{ij}$ and $K_{ji}$ or $\sigma_{ij}$ and $\sigma_{ji}$ to be different. We think that tue current model, rather than representing a real world system faithfully, it help us to understand the behavior of agents in an ideal cooperative scenario. In real world, there are mixed types of interactions including cooperation and competition. For example, in real world, it is expected that some groups of agents cooperate with each other within the group and compete with agents outside the group. We plan to study these more realistic multiagent systems in the future by simulations.

 It is outside the aim of this paper to discuss whether the economic environment determines whether the fair types or the selfish types dominate equilibrium behavior, nor whether cooperation or competition has to be favorized  \cite{fehr1999theory}.  It should be  surely  interesting in further work to adapt the model to very  specific cases,  e.g.,  to  research, sport or other socio-economic activities.  Recall that the time scale can be adapted. Moreover spatial distribution \cite{Damero}, transport costs, and similar economic considerations could introduce new parameters.

\vskip0.5cm {\bf  Acknowledgements}  This paper is part of MA scientific activities in COST Action IS1104,  ``The EU in the new complex geography of economic systems: models,
 tools and policy evaluation",  in COST Action TD1210 `Analyzing the dynamics of information and knowledge landscapes', and   in COST Action  TD1306  ``New Frontiers of Peer Review".

  \clearpage
\begin{center}
\centering 
\begin{table} 
\caption{\label{tab:tabla1}Fixed points and their  $\sigma$ stability interval for the $n=5$ agent case, for various $K<0$.}   
\begin{tabular}{|c|c|c|
|c|c|c|c|c|c|}
\hline

\hline 
 Cases 
&  N$^\circ$ Levels 
& {N$^\circ$ Agents} & $K=-0.0625$ & $K=-0.125$ & $K=-0.25$ & $K=-0.5$ & $K=-1$ \\ \hline 

& & by &   4 overlapping  &   4 overlapping    &   3 overlapping    & only one   &  non stable   \\ 
& &  level  &   stable configurations  &    stable configurations  & stable   configurations  &  stable configuration  &   configuration \\ 
\hline  
I & 1 & 5 &   always stable &   always stable &   never stable &   never stable &     never stable  \\ 
\hline 
II & 2 & 4-1 &  $0.003 \leq \sigma \leq 0.140$ &   $0.004 \leq \sigma \leq 0.387$ &       $0.014 \leq \sigma$ &   never stable   & never stable \\
\hline 
III & 2 & 3-2 &     $0.001 \leq \sigma \leq 0.043$ &     $0.001 \leq \sigma \leq 0.109$ &       $0.001 \leq \sigma \leq 0.4$ &   never stable   & never stable  \\
\hline 
IV & 3 & 3-1-1 &    $0.044 \leq \sigma \leq 0.077$ &     $0.110 \leq \sigma \leq 0.186$ &       $0.410 \leq \sigma\leq 0.630$ &   never stable   & never stable   \\ 
\hline 
V & 3 & 2-2-1 &   $0.003 = \sigma$ &   $0.003 \leq \sigma \leq 0.004$ &       $0.004 \leq \sigma \leq 0.005$ &       $0.007 \leq \sigma \leq 0.009$ &  never stable \\
\hline 
VI & 4 & 2-1-1-1 &     $0.004 \leq \sigma \leq 0.022$ &     $0.005 \leq \sigma \leq 0.049$ &       $0.006 \leq \sigma\leq 0.12$ &       $0.010 \leq \sigma \leq 0.480$   & never stable  \\
\hline 
VII & 5 & 1-1-1-1-1 &   never stable &   never stable &   never stable &   never stable & never stable   \\
\hline 
\end{tabular}
\end{table}

\end{center}

\end{document}